\documentclass[prl,amsmath,twocolumn,amssymb,prl,nobalancelastpage,raggedbottom,superscriptaddress]{revtex4-1}

\usepackage{graphicx}
\usepackage{dcolumn}
\usepackage{bm}
\usepackage[utf8]{inputenc}
\usepackage{lipsum}  
\usepackage{chngpage}
\usepackage{amsmath}
\usepackage{placeins}
\usepackage{xfrac}

\usepackage{amsfonts}%
\usepackage{amssymb}%
\usepackage{feynmp}%
\usepackage{setspace}%
\usepackage{comment}%
\usepackage{dsfont}%
\usepackage{color}%
\usepackage{tikz}
\usepackage{todonotes}
\usetikzlibrary{arrows,calc}

\newcommand{\ket}[1]{\left| #1 \right>} 

\begin{document}

\title{Strong nonreciprocity in modulated resonator chains through synthetic electric and magnetic fields}

\author{Christopher W. Peterson}
\affiliation{Department of Electrical and Computer Engineering}

\author{Wladimir A. Benalcazar}
\affiliation{Department of Physics and Institute for Condensed Matter Theory}

\author{Mao Lin}
\affiliation{Department of Physics and Institute for Condensed Matter Theory}

\author{Taylor L. Hughes}
\affiliation{Department of Physics and Institute for Condensed Matter Theory}

\author{Gaurav Bahl}
\affiliation{Department of Mechanical Science and Engineering, \\ University of Illinois at Urbana-Champaign, Urbana, IL 61801, USA}

\date{\today}

\begin{abstract}
We study nonreciprocity in spatiotemporally modulated 1D resonator chains from the perspective of equivalent 2D resonator arrays with a synthetic dimension and transverse synthetic electric and magnetic fields.
The synthetic fields are respectively related to temporal and spatial modulation of the resonator chain, and we show that their combination can break transmission reciprocity without additional elements.
This nonreciprocal effect is analogous to the Hall effect for charged particles.
We experimentally implement chains of 2 and 3 spatiotemporally modulated resonators and measure over 58 dB of isolation contrast.
\end{abstract}

\maketitle

\begin{figure}[t]
    \centering
    \includegraphics[width = \linewidth]{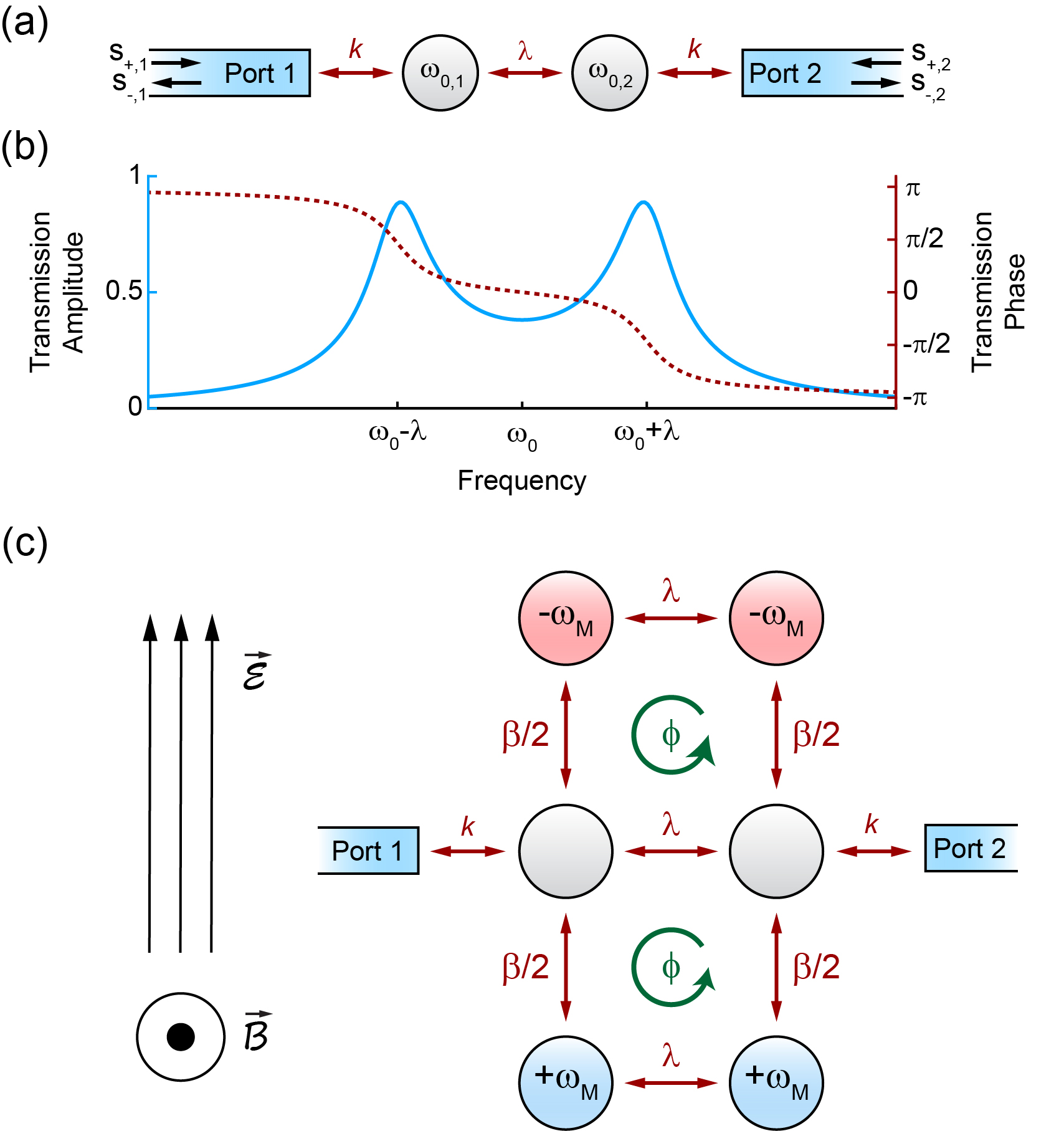}
    \caption{(a) A chain of two coupled resonators coupled to two ports. The resonance frequencies $\omega_{0,n}$ are time-varying as described by Eq. \eqref{mod}.
    (b) Transmission, $t(\omega)$, for the chain of coupled resonators in the absence of modulation. Amplitude (solid blue) and phase (dashed red) are shown separately. The transmission phase is normalized to 0 at $\omega_0$.
    (c) Pictorial representation of $\mathcal{H}$ using synthetic electric field $\vec{\mathcal{E}}$ and magnetic field $\vec{\mathcal{B}}$. The electric field generates a potential of $\mp \omega_M$ on the upper and lower chains respectively, and the magnetic field generates a direction-dependent phase shift of $\phi$ around each closed loop.
    }
    \label{fig_intro}
\end{figure}

Reciprocity is a fundamental property of wave propagation in linear, time-reversal symmetric systems that implies invariance under a spatial inversion of inputs and outputs~\cite{Rayleigh, Carson}, i.e., the scattering matrix $S$ of a reciprocal system is symmetric ($S = S^T$).
Due to this constraint, reciprocal systems cannot provide important functions such as source protection~\cite{What} and directional signal routing~\cite{Hogan}, which are critical to many electromagnetic, optic, and acoustic applications. 
Reciprocity can be broken in linear systems biased with a vector quantity that is odd under time-reversal~\cite{Casimir,EstepIEEE}, e.g., a magnetic field~\cite{Hogan, Lax}, a momentum transfer from spatiotemporal modulation~\cite{Yu,Kamal,Hafezi,Lira,Fang1,Fang2,Estep,Tzuang,Qin,Kim,BanGapMod,Sohn,Peterson}, or a fluid flow~\cite{Fleury}.
In this letter we study nonreciprocity in one-dimensional (1D) chains of coupled photonic resonators with spatiotemporally modulated resonance frequencies. 
We use a synthetic-dimension description of the modulated resonator chains, which can be interpreted as unmodulated 2D resonator arrays with a synthetic frequency dimension~\cite{SynthReview}.
The synthetic dimension holds frequency-shifted copies of the original chain that are equivalent to the sidebands produced by modulation. 
This description is particularly useful because the frequency and phase of the modulation become equivalent to a ``photonic gauge potential" with similar properties to the electromagnetic vector-potential that couples to charged particles~\cite{ArtGauge, SynDim, SynDim2, Walter}.
This gauge potential can generate synthetic electric~\cite{FanBloch} and magnetic~\cite{Tzuang} fields for photons in the resonator array, enabling a rich variety of physical phenomena such as Bloch oscillations~\cite{FanBloch}, topological insulators~\cite{SynDim2,FanWeyl,SingleTopologicalCavity}, and the Aharonov-Bohm effect~\cite{Fang1, Fang2}. 
Reciprocity can be broken in synthetic arrays having a magnetic field, but doing so requires an additional mirror-symmetry-breaking in the frequency dimension \cite{SynDim2} since the synthetic magnetic field is always perpendicular to the plane of the array.
Previous work has relied on additional elements such as filters ~\cite{Fang1, Fang2} or added loss~\cite{SynDim2} to break this symmetry.
Here we introduce a new approach that uses a synthetic electric field to break mirror symmetry in the frequency dimension.
When a synthetic magnetic field that breaks time-reversal symmetry is also present, the combination of the two synthetic fields breaks transmission reciprocity without requiring any additional elements.
This effect is analogous to the Hall effect for charged particles, where perpendicular electric and magnetic fields induce a current in the $\vec{E} \times \vec{B}$ direction \cite{Hall}.
We experimentally verify this concept using short chains of coupled resonators implemented in microwave-frequency microstrip circuits, and observe greater than $58$ dB (approximately six orders of magnitude) of isolation contrast.
We further show that nonreciprocal contrast is maximized when the both synthetic fields are tuned to maximize their respective symmetry-breaking.
As an illustrative case, we first consider a chain of two identical coupled resonators with intrinsic resonance frequencies $\omega_0$, as illustrated in Fig.~\ref{fig_intro}a.
The coupling rate between the resonators is $\lambda,$ and each resonator is also coupled to a port, forming a two-port coupled-cavity waveguide.
The resonance frequency of the resonators is modulated sinusoidally with frequency $\omega_M$ according to
\begin{equation} \label{mod}
\begin{split}
    \omega_{0,1} &= \omega_0 + \beta \cos \big( \omega_M t \big) \\
    \omega_{0,2} &= \omega_0 + \beta \cos \big( \omega_M t + \phi \big).
\end{split}
\end{equation}
The excitation amplitudes $a_{1,2}$ of resonators 1 and 2 (left and right circles in Fig.~\ref{fig_intro}a, respectively) can be collected in a vector $\ket{a(t)}= \begin{bmatrix} a_1(t), a_2(t) \end{bmatrix}^T$. Following Ref.~\onlinecite{CMT}, $\ket{a(t)}$ evolves in time according to
\begin{equation} \label{eq_a}
        \frac{\partial}{\partial t} \ket{a(t)} = (i \Omega_0 + i \Omega_1(t) - \Gamma) \ket{a(t)} + iK^T \ket{s_+(t)},
\end{equation}
where the system parameters are written as the matrices
\begin{equation}
    \Omega_0 = \begin{pmatrix}
    \omega_0 & \lambda \\
    \lambda & \omega_0
    \end{pmatrix}, 
    ~\Gamma =\begin{pmatrix}
    \gamma & 0 \\
    0 & \gamma
    \end{pmatrix},
    ~K = \begin{pmatrix}
    k & 0\\
    0 & k
    \end{pmatrix},\nonumber
\end{equation}
\begin{equation}
    \Omega_1(t) = \begin{pmatrix}
    \beta \cos(\omega_Mt) & 0\\
    0 & \beta \cos(\omega_Mt+\phi) 
    \end{pmatrix}, \nonumber
\end{equation}
and $\ket{s_+(t)}, \ket{s_-(t)}$ are vectors that respectively correspond to the input and output amplitudes at the ports.
The coupling between the ports and resonators is described by the coupling matrix $K$ ($k$ is the coupling constant between a resonator and a port).
The total decay rates of the resonators are described by the matrix $\Gamma$ ($\gamma$ is the decay rate of each resonator), which satisfies $2 \Gamma = K^{\dagger} K + \kappa$ \cite{CMT}. 
The $K^{\dagger} K$ term accounts for the fields decaying into the ports, while the diagonal matrix $\kappa$ accounts for any resistive or radiative losses in each resonator.
The output of the system can be written as
\begin{equation} \label{eq_s}
    \ket{s_-(t)} = \ket{s_+(t)} + iK \ket{a(t)}.
\end{equation}

Since the system is periodic in time with frequency $\omega_M,$ it is convenient to work in the frequency domain. Using the Fourier transform $\ket{a(\omega)}=\int dt \ket{a(t)}e^{-i\omega t}$, in steady-state Eq.~(\ref{eq_a}) becomes
\begin{equation} 
\label{eq_aw}
\begin{split}
    \omega \ket{a(\omega)} & = H_0 \ket{a(\omega)} + K^T\ket{s_+(\omega)} \\ & + B \ket{a(\omega-\omega_M)} + B^{\dagger} \ket{a(\omega+\omega_M)}
     ,
\end{split}
\end{equation}
where $H_0 = \Omega_0 + i\Gamma$ and $$B= \beta / 2 \begin{pmatrix}
1 & 0\\
0 & e^{i\phi}
\end{pmatrix}.$$ 
The applied modulation converts the input signal up and down in frequency such that inputs with a single frequency will generate infinitely many sidebands equally separated by multiples of $\pm\omega_M$. These sidebands are coupled to each other through the $B$ matrix. 
Thus, Eq.~\eqref{eq_aw} is actually a set of infinitely many equations that can be written as \cite{ShirleyFloquet}
\begin{equation} \label{FloquetIn}
    \omega \ket{\alpha(\omega)} = \mathcal{H} \ket{\alpha(\omega)} + \mathcal{K}^T \ket{\sigma_+(\omega)},
\end{equation}
where $\mathcal{K}$ is a block-diagonal matrix where each block is $K$, and $\mathcal{H}$ is the block-tridiagonal matrix
\begin{equation}
    \mathcal{H} = \begin{pmatrix} \ddots & \ddots & 0 & 0 & 0 \\ \ddots &  H_0 - \omega_M I_2 & B & 0 & 0 \\ 0 & B^{\dagger} & H_0 & B & 0 \\ 0 & 0 & B^{\dagger} & H_0 + \omega_M I_2 & \ddots \\ 0 & 0 & 0 & \ddots & \ddots \end{pmatrix}. \nonumber
\end{equation}
The amplitude vectors $\ket{\alpha(\omega)}$ and $\ket{\sigma_{\pm}(\omega)}$ are 
\begin{equation}
     \ket{\alpha(\omega)} = \begin{pmatrix} \vdots \\ \ket{a(\omega + \omega_M)} \\ \ket{a(\omega)} \\ \ket{a(\omega - \omega_M)} \\ \vdots \end{pmatrix}, ~ \ket{\sigma_{\pm}(\omega)} = \begin{pmatrix} \vdots \\ \ket{s_{\pm}(\omega + \omega_M)} \\ \ket{s_{\pm}(\omega)} \\ \ket{s_{\pm}(\omega - \omega_M)} \\ \vdots \end{pmatrix}.
     \nonumber
\end{equation}
The output of the system $\ket{\sigma_-(\omega)}$ can be found through the expression
\begin{equation} \label{FloquetOut}
    \ket{\sigma_-(\omega)} = \ket{\sigma_+(\omega)} + i \mathcal{K} \ket{\alpha(\omega)}.
\end{equation}
The relation between the input and output of the system is given by the expression $\ket{\sigma_-(\omega)} = \mathcal{S}(\omega) \ket{\sigma_+(\omega)}$, where the full scattering matrix $\mathcal{S}(\omega)$ can be found by solving Eqs. \eqref{FloquetIn} and \eqref{FloquetOut}.
However, the linear scattering matrix $S(\omega)$ which only relates inputs and outputs at the same frequency and satisfies $\ket{s_-(\omega)} = S(\omega) \ket{s_+(\omega)}$ is typically more useful, especially if the input is monochromatic $\ket{\sigma_+(\omega)} = \ket{s_+(\omega)}$.
This simpler scattering matrix can be found by using perturbation theory that ignores sideband terms beyond a certain order in the squared modulation amplitude $\beta^2$.
In the following analysis, we aim to illustrate the origin of the linear nonreciprocal effect and therefore solve for $S(\omega)$ by keeping only the first-order sidebands (perturbation order $\beta^2$) \footnote{We keep terms up to the fifth order ($\pm 5 \omega_M$) in all theoretical calculations in this paper.} and neglecting the coupling between the sidebands and the ports.
The transfer function of the unmodulated two-resonator chain described by $H_0$ is $$\tau(\omega)= \frac{k^2 \lambda}{[\gamma + i (\omega-\omega_0)]^2+\lambda^2},$$ as shown in Fig. \ref{fig_intro}b.
The transmission amplitude is symmetric about the center frequency $\omega_0$, but the transmission phase is anti-symmetric --- as we will show, this is vital to breaking reciprocity.
There are two peaks in the transmission amplitude, corresponding to the eigenmode frequencies $\omega_0 \pm \lambda$.
Near these resonant frequencies, the transmission phase is $\approx \mp \pi/2$ relative to the transmission phase at $\omega_0$.
The modulated chain can be interpreted as a time-invariant 2D array with a synthetic dimension arising in frequency space.
This array consists of the original chain and two additional chains for each perturbation order (one additional chain for each sideband).
To first order, the system consists of three unmodulated two-resonator chains separated in frequency, as shown in Fig. \ref{fig_intro}c.
The coupling rate $\beta/2$ between neighboring chains/sidebands is determined by the amplitude of the applied modulation.
We can capture the effects of the modulation frequency and phase by introducing two synthetic fields - an electric field $\vec{\mathcal{E}}$ pointing parallel to the frequency axis, and a magnetic field $\vec{\mathcal{B}}$ pointing out of the 2D plane.
The electric field manifests as a potential gradient of $\omega_M$ between the resonator chains, and is equivalent to the frequency offset of the $H_0$ matrix along the diagonal of $\mathcal{H}$.
The magnetic field produces a magnetic flux that induces a direction-dependent phase shift of $\phi$ in each plaquette \cite{Fang1, Tzuang}, equivalent to the the phase term $e^{i\phi}$ in the $B$ matrix.
For simplicity, and because distance is not well defined in the synthetic dimension, we adopt units where $|\vec{\mathcal{E}}| = \omega_M$ and $|\vec{\mathcal{B}}| = \phi$.

Transmission through the system shown in Fig. \ref{fig_intro}c can be calculated as the sum of transmission through three channels: the central channel with no potential offset, the lower channel with a positive offset $+\omega_M$, and the upper channel with a negative offset $-\omega_M$.
The lower and upper channels both enclose a synthetic magnetic flux which induces an additional direction-dependent phase shift of $\pm \phi$, such that the total transmission, written as a sum of symmetric and anti-symmetric parts, is
\begin{equation} \label{twoparts}
\begin{split}
        S_{21}(\omega) &\approx \tau + \frac{\beta^2}{4} \big( [\tau_+ + \tau_-] \cos(\phi) - i [\tau_+ - \tau_-] \sin(\phi) \big),\\
        S_{12}(\omega) &\approx \tau + \frac{\beta^2}{4} \big( [\tau_+ + \tau_-] \cos(\phi) + i [\tau_+ - \tau_-] \sin(\phi) \big),
\end{split}
\end{equation}
where $\tau = \tau(\omega)$ and $\tau_\pm = \tau(\omega \pm \omega_M)$. 
From Eq. \eqref{twoparts}, we find that the symmetric part of the transmission is identical between $S_{21}$ and $S_{12}$ (i.e., is reciprocal), while the anti-symmetric part differs (i.e., is nonreciprocal).
It is immediately clear that if either synthetic field vanishes the system must be reciprocal, since $\sin(\phi = 0) = 0$ and $\tau_+ = \tau_-$ if $\omega_M = 0$. 
Furthermore, the strongest nonreciprocal response will arise when both $\omega_M$ and $\phi$ are tuned such that transmission is maximally anti-symmetric with respect to the sign of either quantity.
This occurs when the input frequency $\omega = \omega_0$, the synthetic flux $\phi = \pm \pi/2$, and the synthetic potential $\omega_M \approx \lambda$, such that the lower and upper paths are resonant and provide a $\pm \pi/2$ phase shift respectively. 
The resonance of the lower and upper paths also maximizes the amplitudes of $\tau_{\pm}$ and therefore the $\beta^2$ term of Eq. \eqref{twoparts}, further increasing the nonreciprocal contrast.
The effect of the synthetic electric and magnetic fields can be interpreted as a Hall effect for photons.
In the ordinary Hall effect, current flows perpendicular to applied electric and magnetic fields because the combination of fields exerts a force that makes it more favorable for electrons to move in one direction.
Here, the same combination of fields makes it more favorable for photons to move in one direction, leading to transmission nonreciprocity.
This effect is resonantly enhanced in our system, leading to a strong nonreciprocal contrast.

\begin{figure}[t]
    \centering
    \includegraphics[width = \linewidth]{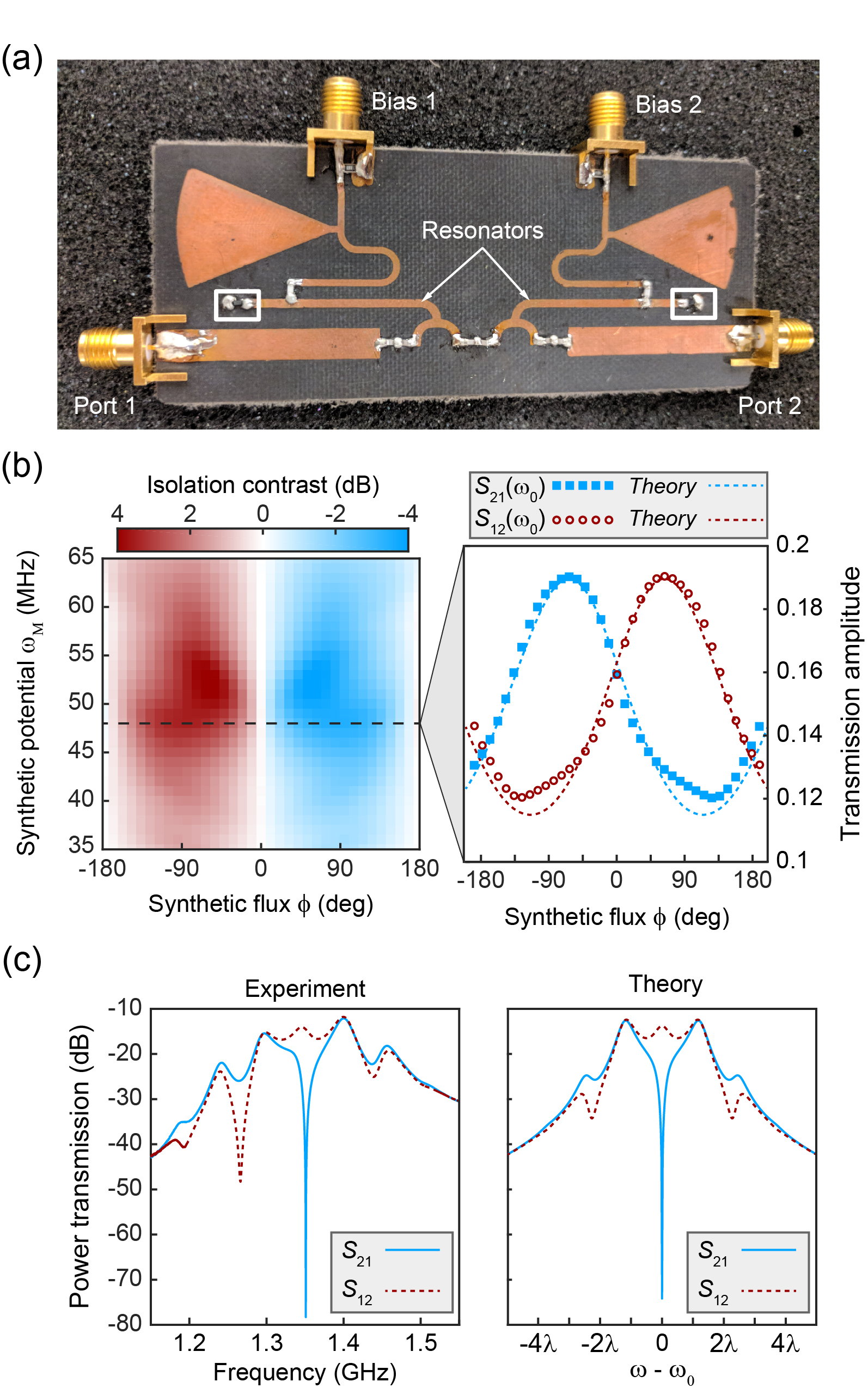}
    \caption{(a) Photograph of the experimental circuit implementing two coupled resonators with variable resonance frequencies. The bias ports are used to apply a voltage to the varactor diodes (white boxes) and thereby modulate the resonant frequency.
    (b, left) Measured isolation contrast at $\omega_0$ as a function of $\omega_M$ and $\phi$. (b, right) Measured forward and backward transmission amplitude at $\omega_M / 2 \pi = 48$ MHz, which corresponds to the dashed line on the left.
    (c) Measured and fitted power transmission for the circuit in (a), with $\omega_M \approx \lambda$, $\phi = \pi/2$, and $\beta$ tuned to minimize forward transmission amplitude.
    }
    \label{fig_2half}
\end{figure}

To test these predictions we implemented a chain of two coupled resonators with modulated resonance frequencies in a microwave circuit using microstrip stub resonators, as pictured in Fig.~\ref{fig_2half}a.
Each resonator has an initial loaded resonance frequency $\omega_0 / 2 \pi \approx 1.35$ GHz and is terminated in a varactor diode that modulates the resonance frequency in response to an applied voltage.
We used external tunable phase shifters to control the phase shift $\phi$ between the sinusoidal voltage biases applied to the resonators. 
We first swept the amplitudes of the synthetic fields (by adjusting the modulation frequency and phase) to find the values that produce the strongest nonreciprocity.
Fig.~\ref{fig_2half}b shows the measured isolation contrast for an input frequency of $\omega_0$ as a function of the synthetic potential $\omega_M \propto \vec{\mathcal{E}}$ and synthetic magnetic flux $\phi \propto \vec{\mathcal{B}}$.
As predicted, the contrast is maximized near specific values of these parameters: $\omega_M / 2 \pi \approx \lambda / 2 \pi \approx 48$~MHz, and $\phi \approx \pi/2$.
The frequency dependence of the circuit components produce an additional asymmetry, not accounted for in our model which has frequency-independent parameters, that shifts the maximum contrast to $\omega_M / 2\pi = 53$~MHz.
The measured transmission amplitude for an input at $\omega_0$ is shown in Fig. \ref{fig_2half}b for $\omega_M / 2\pi = 48$~MHz. 
There is good agreement between the measured transmission and theoretically calculated transmission. 
This experiment also clearly demonstrates that the synthetic electric and magnetic fields work together to produce a strong nonreciprocal response.
The measured transmission with no synthetic flux ($\phi = 0$) is fully reciprocal, and the isolation contrast decreases as the synthetic potential moves away from $\omega_M \approx \lambda \approx 48$ MHz, as expected.
Next we increased the modulation amplitude to minimize the transmission amplitude in the forward direction ($S_{21} \approx 0$) while maximizing isolation contrast.
Figure~\ref{fig_2half}c shows the measured and calculated values of the power transmission, $|S_{12}(\omega)|^2$ and $|S_{21}(\omega)|^2$, under modulation with this critical amplitude.
Here, the modulation frequency has been increased to $\omega_M / 2\pi = 53$~MHz in order to maximize the contrast.
The measured forward transmission approaches zero ($\approx -79$~dB) at $\omega_0$ and measured isolation contrast at $\omega_0$ is $> 64$ dB.
The calculated transmission matches the measured data well and the result is consistent with the prediction of Eq. \eqref{twoparts}.
The synthetic electric and magnetic field interpretation of a modulated resonator chain can be extended to chains of an arbitrary length.
The form of $\mathcal{H}$ remains the same regardless of chain length, only the inner matrices $H_0$ and $B$ change to accommodate more resonators.
For example, in a chain of three resonators
\begin{equation}
    H_0 = \begin{pmatrix}
    \omega_0 + i \gamma & \lambda & 0 \\
    \lambda & \omega_0 + i \gamma & \lambda \\
    0 & \lambda & \omega_0 + i \gamma
    \end{pmatrix}, 
    ~B =\begin{pmatrix}
    1 & 0 & 0\\
    0 & e^{i \phi} & 0\\
    0 & 0 & e^{2 i \phi}
    \end{pmatrix}. \nonumber
\end{equation}
A detailed explanation of the coupled-mode theory for longer modulated chains is provided in the Supplement.
\begin{figure}[t]
    \centering
    \includegraphics[width = \linewidth]{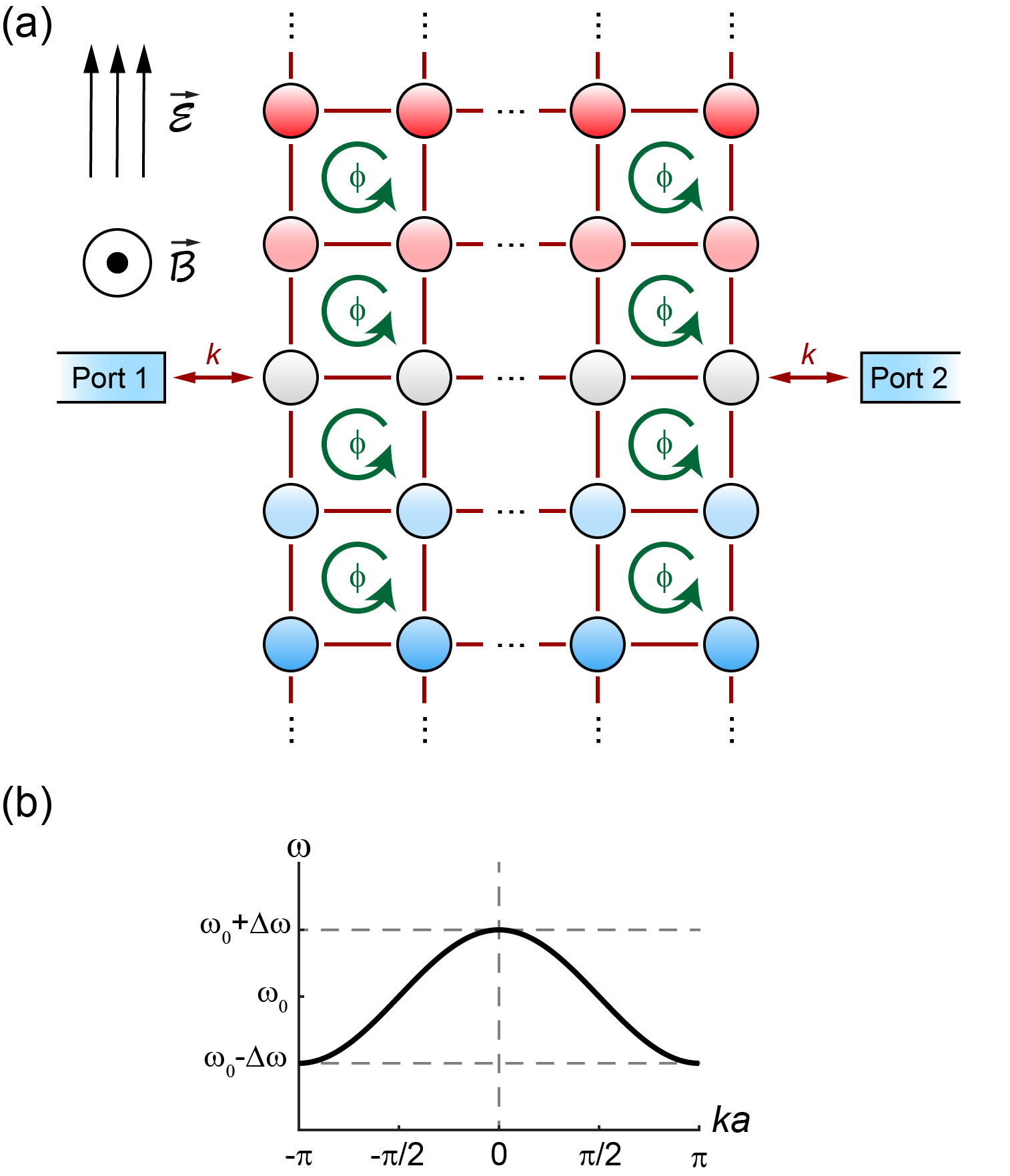}
    \caption{(a) Representation of $\mathcal{H}$ with an arbitrary chain length and number of sidebands. The synthetic electric field induces a potential of $\omega_M$ in the vertical direction and the synthetic magnetic field induces a direction-dependent phase shift of $\phi$ in each plaquette.
    (b) Dispersion relation of an infinite resonator chain with wavevector $k$ and lattice constant $a$. Finite chains with length $N$ have a discrete dispersion of $N$ modes that follows the same trend.}
    \label{fig_longer}
\end{figure}
A graphical representation of $\mathcal{H}$ for a modulated chain with an arbitrary length, and including an arbitrary number of sidebands, is shown in Fig. \ref{fig_longer}a.
Since the number of possible transmission channels increases as the chain length increases, there are multiple mechanisms that can produce strong nonreciprocity in longer chains.
However, since longer chains have an anti-symmetric phase response, the first-order nonreciprocal mechanism that we have identified in the two-resonator chain can be realized in a chain of any length. 
The dispersion relation of an infinite resonator chain is shown in Fig. \ref{fig_longer}b; chains of length $N$ have a discrete dispersion of $N$ modes that follows the same trend.
The outermost eigenmodes of resonator chains always follow the pattern found in the two-resonator chain: the phase difference between adjacent resonators ($ka$ in Fig. \ref{fig_longer}b), relative to the phase difference at $\omega_0$, is $+\pi/2$ for the lowest-frequency mode and $-\pi/2$ for the highest mode.
Hence, strong transmission nonreciprocity occurs for inputs at the center frequency $\omega_0$ when the synthetic flux $\phi = \pi/2$ and the synthetic potential $\omega_M = \Delta \omega$, where $\Delta \omega$ is the frequency separation between the outermost eigenmodes and the center frequency. 
In the shortest case of two resonators $\Delta \omega = \lambda$, and as the chain length increases $\Delta \omega \rightarrow 2 \lambda$.
A theoretical analysis of how this mechanism works in a three-resonator chain is provided in the Supplementary Information. 
We implemented a chain of three modulated resonators using three microstrip resonators with voltage-controlled resonance frequencies (photo in Supplement).
Here, each resonator has an initial loaded resonance frequency $\omega_0 \approx 1.32$~GHz.
The measured and calculated transmission spectra are shown in Fig.~\ref{fig_3full}a for $\phi = \pi/2$ and $\omega_M / 2 \pi = \Delta \omega / 2 \pi \approx 141$~MHz, where the modulation amplitude is tuned to minimize the forward transmission amplitude.
As in the two-resonator chain, the measured forward transmission near $\omega_0$ approaches zero ($\approx -80$~dB), and there is strong nonreciprocal contrast ($\approx 59$~dB).
\begin{figure}[t]
    \centering
    \includegraphics[width = \linewidth]{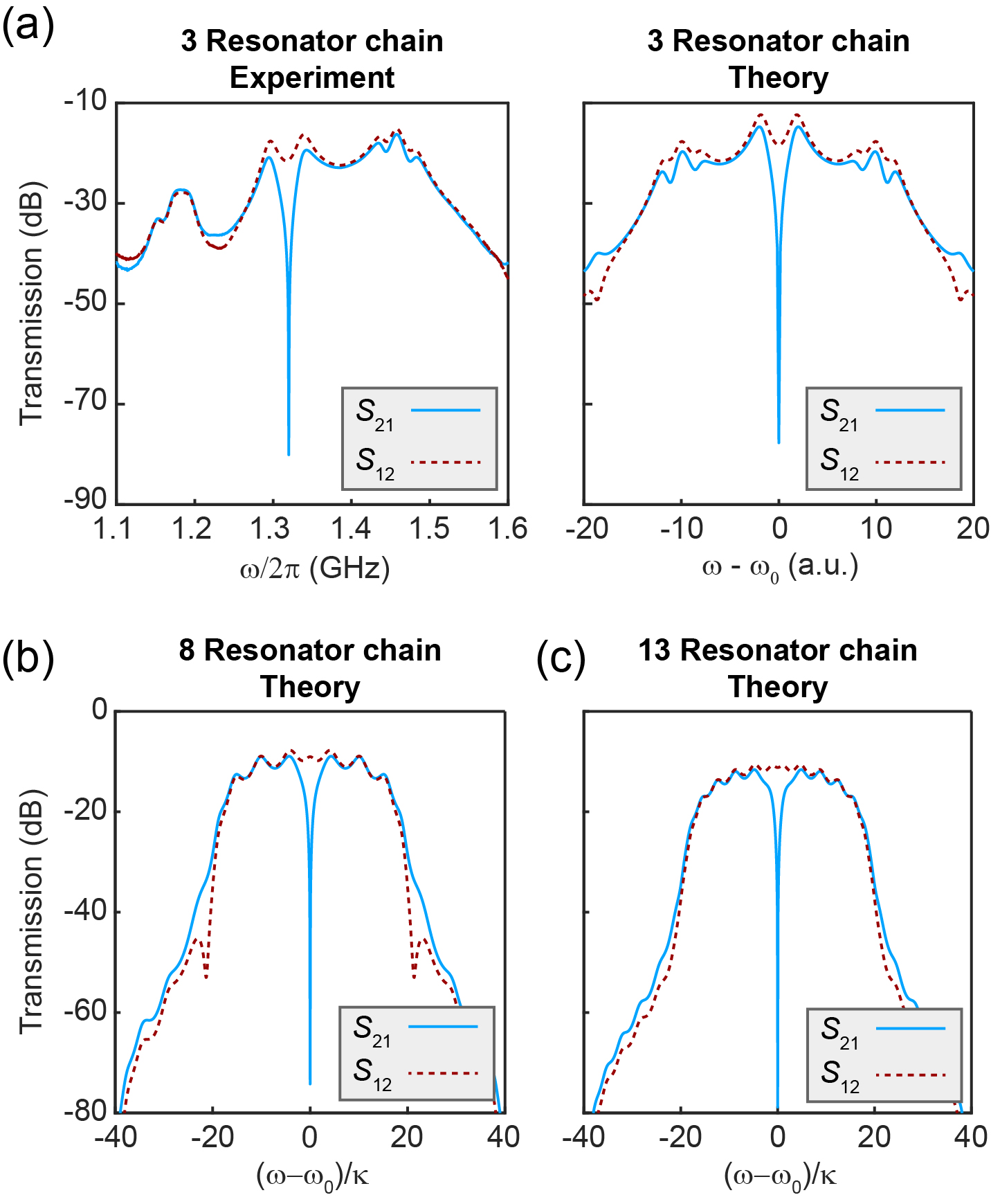}
    \caption{(a) Measured and fitted power transmission for the three-resonator chain, with $\omega_M = \Delta \omega$, $\phi = \pi/2$, and $\beta$ tuned to minimize forward transmission amplitude.
    (b) Calculated transmission for an 8 resonator chain under modulation with $\omega_M = \Delta \omega$, $\phi = \pi/2$, and $\beta$ tuned to minimize forward transmission amplitude.
    (c) Calculated transmission for a 13 resonator chain under modulation with $\omega_M = \Delta \omega$, $\phi = \pi/2$, and $\beta$ tuned to minimize forward transmission amplitude.
    }
    \label{fig_3full}
\end{figure}
We also simulated longer chains using the coupled-mode theory model with intrinsic resonator linewidth $\kappa$, $\lambda = 10 \kappa$, $k = 2 \kappa$, $\phi = \pi/2$, and keeping sideband terms up to $\pm 5\omega_M$.
As the chain length increases, the number of eigenmodes of the chain increases proportionally, eventually forming a flat passband around $\omega_0$ as the modes overlap due to their finite linewidth.
We plot the forward ($S_{21}$) and backward ($S_{12}$) transmission spectra for chains of 8 and 13 resonators under modulation with $\phi = \pi/2$, $\omega_M = \Delta \omega$, and $\beta$ tuned to minimize $S_{21}(\omega_0$) in Fig.~\ref{fig_3full}b,c.
Despite the increased number of eigenmodes forming a passband instead of discrete resonances, the major nonreciprocal feature remains the large nonreciprocal dip near $\omega_0$.
Furthermore, the spectrum is approximately reciprocal between the central frequency and the band edges, indicating that the main nonreciprocal mechanism is the first-order process related to the anti-symmetric eigenmodes.
The synthetic field interpretation of spatiotemporal modulation presented here can be applied to a wide variety of nonreciprocal systems \cite{Yu,Lira,Fang1,Fang2,Estep,Tzuang,Peterson}.
Additionally, the nonreciprocal mechanism we proposed is general and can be realized without additional filters in any system of coupled resonators with the appropriate anti-symmetric modes, including chains of coupled photonic or electronic crystal defects.
Since resonance frequency modulation is practical across a variety of resonator types \cite{Xu, Reed, Estep, Mert}, the proposed method for generating strong nonreciprocity can be implemented across domains, in optical, microwave, or mechanical resonators.

\vspace{12pt}

\section*{Acknowledgements}

We thank Prof. Jennifer T. Bernhard for access to the resources at the UIUC Electromagnetics Laboratory. 
This project was supported by the US National Science Foundation (NSF) grants EFMA-1627184 (EFRI) and DMR-1351895, the US Office of Naval Research Director for Research Early Career Grant, and an NSF Graduate Research Fellowship for CWP.

\newpage

\newcommand{\beginsupplement}{%
        \setcounter{table}{0}
        \renewcommand{\thetable}{S\arabic{table}}%
        \setcounter{figure}{0}
        \renewcommand{\thefigure}{S\arabic{figure}}%
        \setcounter{equation}{0}
        \renewcommand{\theequation}{S\arabic{equation}}
        \setcounter{section}{0}
        \renewcommand{\thesection}{S\arabic{section}}%
}

\beginsupplement

\begin{center}
\LARGE{Supplementary Information} \\
\end{center}

\vspace{5mm}

\section{S-matrix for general modulated N-resonator system}

In the main manuscript, we have derived the Hamiltonian $\mathcal{H}$ and equations of motion for a spatiotemporally modulated two-resonator chain.
In this section, we generalize the calculation for a chain of length $N$ and also calculate the scattering matrix, including the effects of arbitrarily high perturbation order (number of sidebands).
We recall the Fourier transformed equation of motion as
\begin{subequations} \label{eq:EqFrequency_2}
\begin{equation} 
    \begin{split}
        \omega \ket{a(\omega)} &= H_0 \ket{a(\omega)} + K^T \ket{s_+(\omega)} \\ &+ B \ket{a(\omega - \omega_M)} + B^{\dagger} \ket{a(\omega + \omega_M)}, \label{eq:EqFrequency_2_1}
    \end{split}
\end{equation}
\begin{equation} \label{eq:EqFrequency_2_2}
    \ket{s_-(\omega)} = \ket{s_+(\omega)}+iK\ket{a(\omega)},
\end{equation}
\end{subequations}
where $H_0$ and $B$ are $N\times N$ matrices describing the coupling between the central and side bands. 
$K$ is a $2\times N$ matrix that couples the resonant modes to the external ports. 
For example, in a chain with three resonators the matrices that define the system are
\begin{equation} \label{three1}
H_0 = \begin{pmatrix} \omega_0 & \lambda & 0 \\ \lambda & \omega_0 & \lambda \\ 0 & \lambda & \omega_0 \end{pmatrix} + i \begin{pmatrix} \gamma_1 & 0 & 0 \\ 0 & \gamma_2 & 0 \\ 0 & 0 & \gamma_3 \end{pmatrix}~,
\end{equation}
\begin{equation}
    B =\frac{\beta}{2}\begin{pmatrix} 1 & 0 & 0 \\ 0 & e^{i\phi} & 0 \\ 0 & 0 & e^{2i\phi} \end{pmatrix}~,
\end{equation}
\begin{equation} \label{three2}
    K = \begin{pmatrix} k_1 & 0 & 0 \\ 0  & 0 & k_2 \end{pmatrix}~.
\end{equation}
We note that the specific forms of these matrices do not impact the overall discussion, and Eqs. \ref{eq:EqFrequency_2} are applicable for a generic $N$-resonator system. 
From Eqs.~\eqref{eq:EqFrequency_2} it is clear that, given a monochromatic input, the output will consist of multiple frequencies since the applied modulation converts the input signal up and down in frequency, creating sidebands separated by $\omega_M$. To simplify the notation of these sidebands we label the Hamiltonian and amplitude vectors as
\begin{eqnarray}\begin{aligned}
H_p = H_0 - p \omega_M I  , \\
\ket{a(\omega+p~\omega_M)} &\equiv \ket{a_p} , \\
\ket{s_-(\omega+p~\omega_M)} &\equiv \ket{s_{-,p}} ,
\end{aligned}\end{eqnarray}
where $p\in{\bf Z}$. Thus, Eqs.~\eqref{eq:EqFrequency_2} can be written as
\begin{subequations}
\begin{equation} 
    \begin{split}
        \omega \ket{a_p} &= H_0 \ket{a_p} + K^T \ket{s_{+,p}} \\ &+ B \ket{a_{p-1}} + B^{\dagger} \ket{a_{p+1}}, \label{subeq:EOM_omega_Z_1}
    \end{split}
\end{equation}
\begin{equation}
    \ket{s_{-,p}} = \ket{s_+(\omega)}+K\ket{a_p}, \label{subeq:EOM_omega_Z_2}
\end{equation}
\end{subequations}
We keep the notation for $\ket{s_+(\omega)}$ because the input signal is monochromatic. Likewise, the S-matrix that describes scattering from the input frequency to the sidebands can be defined as
\begin{eqnarray}\begin{aligned}
\label{eq:def_S_matrix}
\ket{s_{-,p}}  = S_p\ket{s_+(\omega)}
\end{aligned}\end{eqnarray}
where $S_p \equiv S(\omega+p~\omega_M)$. Comparing Eq.~\eqref{eq:def_S_matrix}, to Eq.~\eqref{subeq:EOM_omega_Z_2}, we aim to solve $\ket{a_p}$ in terms of $\ket{s_+(\omega)}$. 
To do that, we notice Eq.~\eqref{subeq:EOM_omega_Z_1} defines an infinitely extended recursion relating $\ket{a_p}$ to $\ket{a_{p+1}}$ and $\ket{a_{p-1}}$.
If the recursion relation is truncated by keeping only the contributions up to $\pm P$, then the recursive equations can be written as the following block tridiagonal form

\begin{widetext}
\begin{equation}
\label{eq:Recursion_block_form}
\omega \begin{bmatrix}
\ket{a_{P}} \\
\ket{a_{P-1}} \\
\vdots \\
\ket{a_0} \\
\vdots \\
\ket{a_{-(P-1)}} \\
\ket{a_{-P}}
\end{bmatrix}
= 
\begin{bmatrix}
H_P & B \\ 
B^{\dagger} & H_{P-1} & B \\
& \ddots & \ddots & \ddots & \\
& & B^{\dagger} & H_{0} & B \\
& & & \ddots & \ddots & \ddots & \\
& & & & B^{\dagger} & H_{-(P-1)} & B \\
& & & & & B^{\dagger} & H_{-P}  \\
\end{bmatrix}
\begin{bmatrix}
\ket{a_{P}} \\
\ket{a_{P-1}} \\
\vdots \\
\ket{a_0} \\
\vdots \\
\ket{a_{-(P-1)}} \\
\ket{a_{-P}}
\end{bmatrix}
+ 
\begin{bmatrix}
\\
\\
\\[0.7em]
K^T\ket{s_+(\omega)}\\
\\[-0.7em]
\\ 
\\
\\
\end{bmatrix}.
\end{equation}

The inverse of a block tridiagonal matrix is well known in the literature, see for example Ref.~\onlinecite{bowden_direct_1989}. Solving for the field at the input frequency, $\ket{a_0}$, we find
\begin{eqnarray}\begin{aligned}
\label{eq:sol_a_0}
\ket{a_0} = \left\{[\omega I-H_0]+O_{-P}+\tilde{O}_{P}\right\}^{-1}K^T\ket{s_+(\omega)},
\end{aligned}\end{eqnarray}
where 
\begin{eqnarray}\begin{aligned}
\label{eq:O_matrices}
O_{-P} &= B([\omega I - H_{-1}] + B( [\omega I - H_{-2}] +  B ([\omega I - H_{-3}] + ... +B ( [\omega I - H_{-(P-1)}] \\ & + B [\omega I -  H_{-P}]^{-1}B^* )^{-1} B^{\dagger} ... )^{-1} B^{\dagger} )^{-1} B^{\dagger} )^{-1} B^{\dagger},\\
\tilde{O}_{P} &= B^{\dagger}([\omega I - H_1] + B^{\dagger}( [\omega I - H_2] +  B^{\dagger} ([\omega I - H_3] + ... + B^{\dagger} ( [\omega I - H_{P-1}] \\ & + B^{\dagger}[\omega I - H_P]^{-1}B )^{-1} B ... )^{-1} B )^{-1} B )^{-1} B.
\end{aligned}\end{eqnarray}
\end{widetext}
From Eq.~\eqref{eq:sol_a_0}, the scattering matrix for the central frequency can be written
\begin{eqnarray}\begin{aligned}
\label{eq:S_matrix_n}
S_0&=I+K\left\{[\omega I - H_0]+O_{-P}+\tilde{O}_{P}\right\}^{-1}K^T.
\end{aligned}\end{eqnarray}

\section{Scattering analysis for the modulated 3-resonator system}

\begin{figure}[ht]
    \centering
    \includegraphics[width = \linewidth]{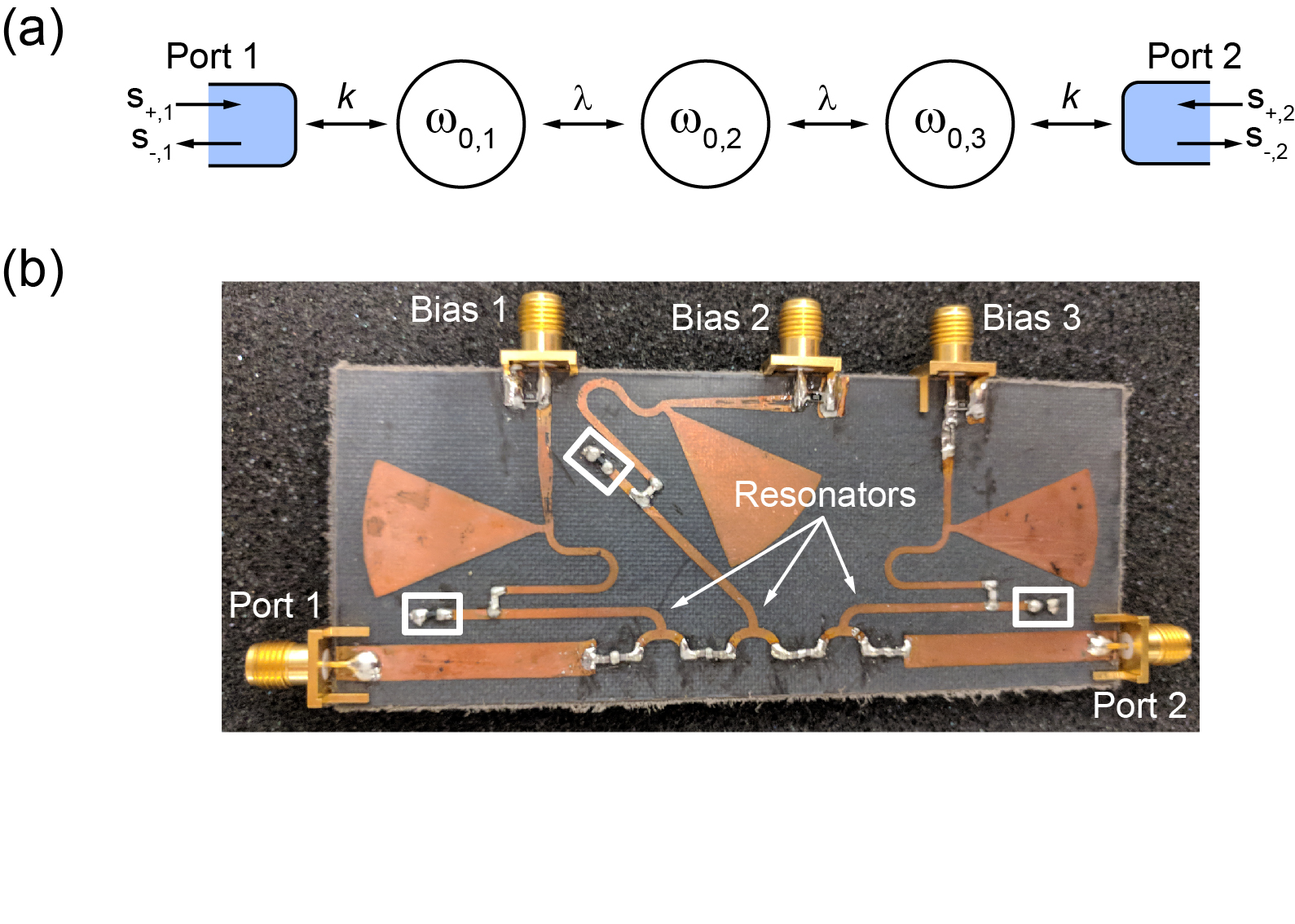}
    \caption{(a) A chain of three coupled resonators coupled to two ports. The resonance frequencies $\omega_{0,n}$ are time-varying as described in the main manuscript.
    (b) Photograph of the experimental circuit implementing three coupled resonators with variable resonance frequencies. The bias ports are used to apply a voltage to the varactor diodes (white boxes) and thereby modulate the resonant frequency.}
    \label{fig_3i}
\end{figure}

\begin{figure}[th]
    \centering
    \includegraphics[width = \linewidth]{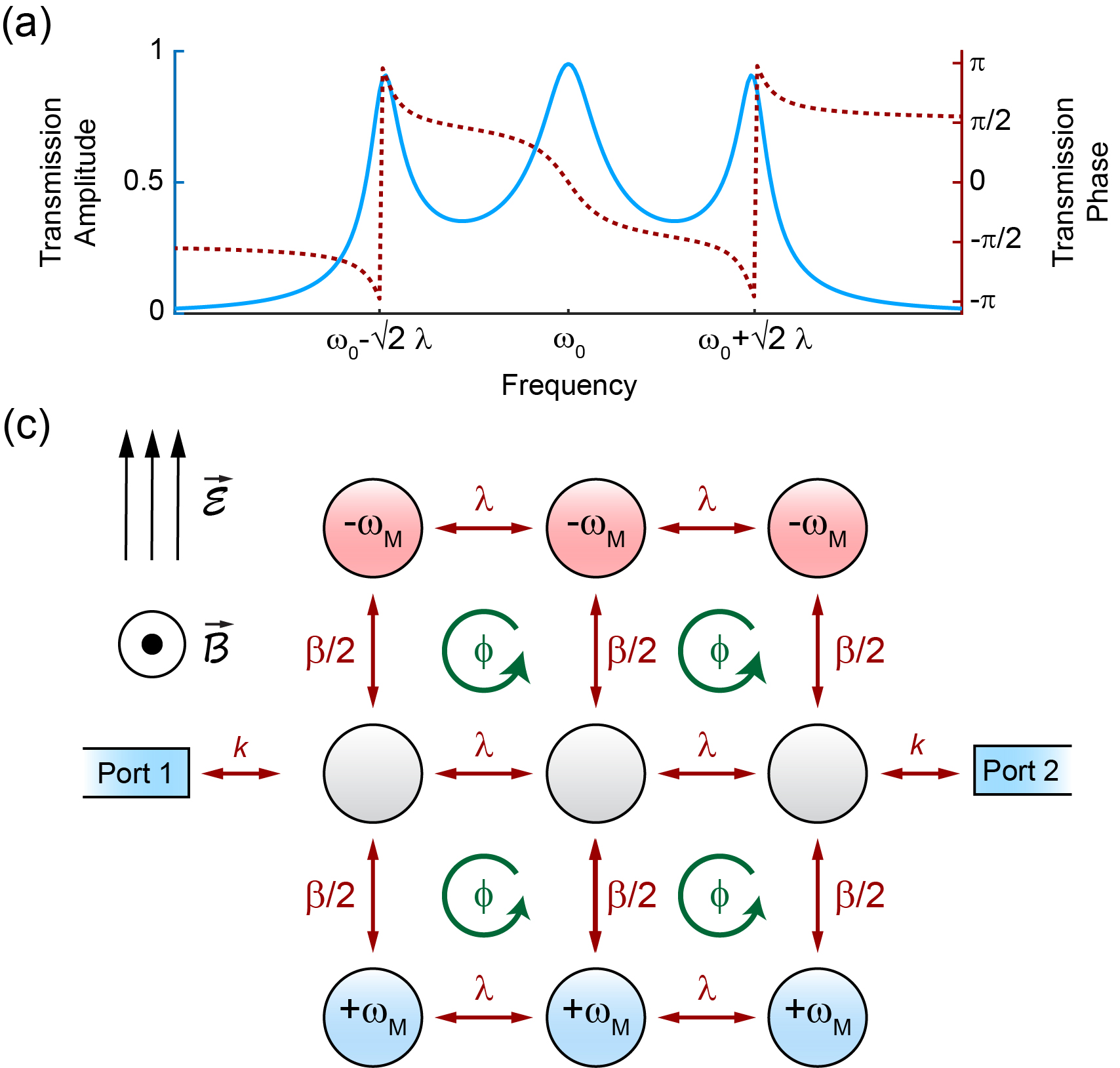}
    \caption{(a) Amplitude (solid blue) and phase (dashed red) of $S_{21}(\omega)$ for the chain of three coupled resonators in the absence of modulation (Eq.~\eqref{eq:response_unmodulated_3_resonator}). 
    (b) Representation of $\mathcal{H}$ for a three-resonator chain with synthetic electric and magnetic fields.}
    \label{fig_3x}
\end{figure}

In this section we provide an intuitive picture of the nonreciprocal behavior of a chain with three resonators (Fig. \ref{fig_3i}a) as the effect of multiple interfering paths. 
This discussion is especially relevant to the experimental results presented in section S1 for the three-resonator chain, and additionally shows how the multiple paths picture of the two-resonator chain can be extended to even longer chains.
The experimental circuit with three resonators is shown in Fig. \ref{fig_3i}b.
We first consider the unmodulated three-resonator chain, which has the following reciprocal response (assuming all resonators have the same decay rate $\gamma$)
\begin{equation} \label{eq:response_unmodulated_3_resonator}
\begin{split}
        S_{21}(\omega) &= S_{12}(\omega) \\& = \frac{k_1k_2\lambda^2}{[\gamma+i(\omega-\bar{\omega})]^3+2\lambda^2[\gamma+i(\omega-\bar{\omega})]} \\
\end{split}
\end{equation}
as shown in Fig.~\ref{fig_3x}a. 
The first-order synthetic lattice representation of this system consists of three unmodulated three-resonator chains that are coupled at each resonator (Fig. \ref{fig_3x}b).
Transmission through this lattice cannot be approximated in the same way as in the two-resonator case (by summing the transmission through each chain with a phase factor) because the chains are coupled at multiple points.
However, knowledge of the eigenmodes of the three-resonator chain allows us to make an approximation that captures the essential aspects of the nonreciprocal mechanism.
The three resonator chain has eigenmodes $|a^\pm\rangle=[1,\pm\sqrt{2},1]^T$, and $|a^0\rangle=[-1,0,1]^T$, which correspond to eigenvalues $\omega=\omega_0\pm \sqrt{2} \lambda$ and $\omega = \omega_0$ respectively.
As in the two-resonator chain, in the lowest frequency mode $|a^-\rangle$ each resonator is excited $\pi$ out-of-phase with its neighbors and in the highest frequency mode $|a^+\rangle$ all resonators are excited in-phase.
At $\omega_0$ there is a $\pi/2$ phase shift between adjacent resonators, which due to destructive interference results in a mode ($|a^0\rangle$) where the middle resonator is not excited.
The basic odd-symmetric structure that characterizes these modes is the same as the two-resonator chain, although here there is a resonant mode at $\omega_0$.

Having understood the unmodulated system, we are ready to consider the modulated system as described in Eqs.~\eqref{three1}-\eqref{three2}. 
Examining Fig.~\ref{fig_3x}b, we find seven paths between the ports that are first-order in $\beta^2$.
There is one path (through the central chain) that encompasses no synthetic flux, two that encompass $2 \phi$ flux, and four that encompass $\phi$ flux.
If $\omega = \omega_0$ and $\omega_M \approx \sqrt{2} \lambda$ such that all paths are resonant and have approximately the same amplitude, we can sum these seven paths (including the phase shift between resonators from the eigenmode analysis) to find that the overall transmission
\begin{equation}
\begin{split} \label{r3_t}
    S_{21}(\omega) &\propto 1 + \frac{\beta^2}{2} \cos(2 \phi) + \beta^2 \sin(\phi), \\
    S_{12}(\omega) &\propto 1 + \frac{\beta^2}{2} \cos(2 \phi) - \beta^2 \sin(\phi).
\end{split}
\end{equation}
Equation \ref{r3_t} reveals that the paths encompassing $2 \phi$ flux are reciprocal, but the paths encompassing $\phi$ flux are not. These paths with $\phi$ flux are the similar to those we identified in the main manuscript for the two resonator chain, but pass through an additional resonator in the central chain. 
We note that the $\cos(2 \phi)$ term that appears in Eq. \ref{r3_t} is related to the normal mode splitting shown in the main manuscript Fig. 3a, which lowers the transmission at $\omega_0$ in the high-transmission direction.

\end{document}